\documentstyle[12pt]{article}
\begin{document}

\author{A. Yu. Segal{\thanks {e-mail: segal@phys.tsu.tomsk.su}} ${}$ and A. A. Sharapov}
\title{Coherent (spin-)tensor fields on D=4 anti-de Sitter space}
\date{{\it {\ Department of Physics, Tomsk State University, Tomsk 634050, Russia}}}
\maketitle

\begin{abstract}
The coherent states associated to the discrete series representations $%
D(E_o,s)$ of $SO(3,2)$ are constructed in terms of (spin-)tensor fields on $%
D=4$ anti-de Sitter space. For $E_o>s+5$ the linear space ${\cal H}_{E_o,s}$
spanned by these states is proved to carry the unitary irreducible
representation $D(E_o,s)$. The $SO(3,2) $-covariant generalized Fourier
transform in this space is exhibited. The quasiclassical properties of the
coherent states are analyzed. In particular, these states are shown to be
localized on the time-like geodesics of anti-de Sitter space.
\end{abstract}

\section{Introduction}

The $D=4$ anti-de Sitter (AdS) space is the maximally symmetrical solution
of the Einstein equations with the negative cosmological constant. This
space arises as a consistent vacuum solution in various modern field
theories coupled to gravity. Among them there are the extended $D=4$
supergravities \cite{SG}, supersymmetric theories of the Kaluza-Klein type
\cite{KK}, interacting theories of the higher-spin massless fields by
Fradkin and Vasiliev \cite{FV},\cite{V}.

The symmetry Lie algebra of the AdS space, $so(3,2)$, has unitary
representations with bounded energy that is crucial for the existence of the
particle interpretation, it is possible also to introduce the notions of
massive and massless particles \cite{I}-\cite{Evans}. Besides, there exist
two remarkable `singleton' representations \cite{DS} having no direct analog
in the flat space
and which were conjectured by Fronsdal and Flato to play a role of `preons'
for the massive and massless particles \cite{Singletons}.

All this indicates that, like the Minkowski space, the AdS space may play an
important role for the particle physics and deserves a serious study.

The essential ingredient of a perturbative field theory is a construction of
the complete basis of solutions for linearized field equations. In this
paper we construct the explicit realizations for the Hilbert spaces of
solutions for the free relativistic wave equations corresponding to the
massive unitary irreducible representations (UIR) of the AdS group. The
absence of the simple analog of the Fourier transform, as distinguished from
Minkowski space case, makes this task not so evident. Our approach to the
problem is based on the use of {\it coherent states method} \cite{Perelomov}%
, that allows, in particular, to present a general solution for the
equations as well as provide it with the transparent quasiclassical
interpretation.

We will proceed as follows. In Sec. 2 we recall some basic facts about the
geometry of AdS space and the positive energy representations of AdS group.
Sec. 3 is devoted to the derivation of families of the coherent states (CS)
associated with the massive representations of the AdS group embedded into
the space of irreducible (spin-)tensor fields on AdS space. In Sec. 4 we
introduce the reproducing kernel associated to the family of CS which is the
central object for the proof of unitarity and irreducibility of the
representations in the linear spaces spanned by the CS set. In Sec. 4 we
discuss the quasiclassical properties of the CS family. In Conclusion we
summarize the results and discuss some perspectives. Appendix contains basic
formulas of $SO(3,2)$-spinor formalism.

\section{Preliminaries}

The four-dimensional AdS space can be realized as a one-sheet hyperboloid
embedded into five-dimensional pseudoeuclidean space ${\bf R}^{3,2}$ with
coordinates $\{y^A\}$, $A=5,0,1,2,3$. The embedding is set by the equation
\begin{equation}
\label{p0}\eta _{AB}y^Ay^B=-R^2\quad \quad ,\qquad \quad diag\quad \eta
_{AB}=(--+++)
\end{equation}
where ${\cal R}=-12R^{-2}$ is the scalar curvature of AdS space. The space
has the topology of ${\bf R}^3\times S^1$ and can be globally parametrized
by intrinsic coordinates $(t,\vec y)$ defined by the rule
\begin{equation}
\label{1p}y^0+iy^5=e^{it}Y\;,\;\;\;Y=\sqrt{R^2+\vec y{}^2}\;,
\end{equation}
$$
0\leq t<2\pi
$$
The induced metric is
\begin{equation}
\label{metric}ds^2=-Y^2dt^2+dy_idy^i-Y^{-2}y_idy^iy_jdy^j
\end{equation}

The isometric embedding (\ref{p0}) allows one to give an extremely simple
geometrical description for the variety of all geodesics of AdS space.
Namely, each geodesic may be visualized as a line of intersection of some
two-plane in ${\bf R}^{3,2}$, passing through the origin, with the surface
of hyperboloid. Any two-plane, in turn, can be spanned by a pair of
orthogonal nonzero five-vectors $u$ and $v$. In what follows we will be
interested only in the time-like geodesics. In this case one may assume that
$u^2=v^2=1$. It is convenient to combine the introduced vectors into the
complex one $p=u+iv$ satisfying the relations
\begin{equation}
\label{p}p^2=0\;\;,\;\;\;(p,\bar p)=-2
\end{equation}
Then the $SO(2)$-rotation of the orthonormal frame $(u,v)$ in the respective
two-plane corresponds to the change of $p$ by a phase factor: $p\rightarrow
e^{i\varphi }p$, and thus the variety of all oriented two-planes is in
one-to-one correspondence with the points of quotient space defined by the
eqs.(\ref{p}) and the equivalence relation
\begin{equation}
\label{pp}p\sim e^{i\phi }p\;\;,\;\;\;\forall \varphi \in {\bf R}
\end{equation}
In order to fix an orientation of the two-planes one can impose the
additional condition~\footnote{%
Actually, in view of (\ref{p}) the absolute value of this expression is
always greater then or equals 1.}
\begin{equation}
\label{ppp}i(p_0\bar p_5-\bar p_0p_5)>0
\end{equation}
The orientation of two-planes induces an orientation on the respective
time-like geodesics that may be thought of as the choice of the time
direction. The inverse orientation results from the complex conjugation: $%
p\rightarrow \bar p$.

The AdS group $SO(3,2)^{\uparrow }\sim Sp(4)/\{\pm 1\}$ is the connected
component of the identity of the group of all pseudoorthogonal
transformations of ${\bf R}^{3,2}.$ It preserves the orientation of
`time-like' two-planes (\ref{p}). The infinitesimal transformations of AdS
group on hyperboloid (\ref{p0}) are generated by ten vector fields
\begin{equation}
\label{5p}{\cal L}_{AB}=y_A\partial _B-y_B\partial _A
\end{equation}
The corresponding Hermitean generators $L_{AB}$ of $so(3,2)$ (in the above
realization $L_{AB}=i{\cal L}_{AB}$) form the algebra
\begin{equation}
\label{p2}[L_{AB},L_{CD}]=i(\eta _{AC}L_{DB}+\eta _{BD}L_{CA}-\eta
_{BC}L_{DA}-\eta _{AD}L_{CB})
\end{equation}
It is easily seen that the AdS group acts transitively on the set of
time-like geodesics: using a proper $SO(3,2)^{\uparrow }$-transformation any
vector $p$ can be transferred to the form $\stackrel{\circ }{p}=(i,1,0,0,0)$%
. The stability subgroup of $\stackrel{\circ }{p}$ coincides with $SO(3)$.
Taking into account the equivalence rel.(\ref{pp}) one concludes that the
set of time-like geodesics is isomorphic to the quotient space
\begin{equation}
\label{7p}{\cal F}=\frac{SO(3,2)^{\uparrow }}{SO(3)\times SO(2)}\sim {\bf R}%
^6
\end{equation}
and eqs.(\ref{p}-\ref{ppp}) present its covariant realization.

In the coordinates (\ref{1p}), the shift $t\rightarrow t+\epsilon $ is the
symmetry transformation generated by the unique time-like Killing vector $%
{\cal L}_{05}$. Therefore, the coordinate $t$ can be naturally identified
with the AdS time, while ${\cal L}_{05}$ may be regarded as the energy
operator.

The positive-energy UIR of $SO(3,2)^{\uparrow }\sim Sp(4)/\{\pm 1\}$ denoted
as $D(E_o,s)$ are classified by two quantum numbers: minimal energy $E_o$
and spin $s$, $s=0,1/2,1,...$. The unitarity requires \cite{Evans}
\begin{equation}
\label{9p}E_o=s+\frac 12\;,\;\;\;s=0,\frac 12\;;\quad E_o\geq
s+1\;,\;\;\;s\geq 1\;.
\end{equation}
$D(1/2,0)$ and $D(3/2,1/2)$ correspond to Dirac singletons \cite
{DS,Singletons}; the representations $D(2,0)$ and $D(s+1,s)$, for any $s$,
describe massless particles; finally, the massive particles are associated
with the choice $E_o>s+1$.

All these representations are of the lowest weight type and, as a result,
they have a quite simple structure. The lowest weight of $D(E_o,s)$ (ground
state) is defined as the eigenstate for the energy and the
third-projection-of-spin operators \refstepcounter{equation}\label{lw}
$$
L_{05}|E_o,s\rangle =E_o|E_o,s\rangle \;\;,\;\;\;\;L_{12}|E_o,s\rangle
=s|E_o,s\rangle \eqno{(\theequation.a)}
$$
annihilated by the corresponding lowering operators%
$$
E_j^{-}|E_o,s\rangle =0\;\;,\;\;\;\;S^{-}|E_o,s\rangle =0\;\;,\;\;\;\;j=1,2,3%
\eqno{(\theequation.b)}
$$
where%
$$
\begin{array}{c}
E_j^{\pm }=(L_{0j}\pm iL_{5j})\;\;,\;\;\;\;S^{\pm }=(L_{13}\pm iL_{32}) \\
\\
\lbrack L_{05},E_j^{\pm }]=\pm E_j^{\pm }\;\;,\;\;\;[L_{12},S^{\pm }]=\pm
S^{\pm }
\end{array}
$$
It is also required for the subspace of states with the minimal energy $E_o$%
, denoted as $|\Lambda \rangle $, to carry the irreducible spin-$s$
representation of the rotation subgroup $SO(3)\subset SO(3,2)^{\uparrow }$.

In this paper we will be interested in the massive representations. In this
case the full basis $B$ of the representations $D(E_o,s)$ may be obtained by
the successive action on $|\Lambda \rangle $ by the rising operators $%
E_j^{+} $:
\begin{equation}
B=\left\{ |\Lambda \rangle ,\quad E_i^{+}|\Lambda \rangle ,\quad
E_i^{+}E_j^{+}|\Lambda \rangle ,\quad ...\;\right\}
\end{equation}

These representations can be also characterized by the eigenvalues of the
quadratic and quartic Casimir operators associated to the $so(3,2)$ algebra
\cite{Evans}
\begin{equation}
\label{cas}
\begin{array}{c}
C_2=\frac 12L_{AB}L^{AB}=E_o(E_o+3)+s(s+1) \\
\\
C_4=W_AW^A=-s(s+1)(E_o-1)(E_o-2) \\
\\
W_A=\frac 18\epsilon _{ABCDE}L^{BC}L^{DE}
\end{array}
\end{equation}

With the given ground state (\ref{lw}) one may associate the family of
coherent states forming the overcomlete basis in the space of $D(E_o,s)$ and
defined by the rule
\begin{equation}
\label{p6}|g\rangle =g|E_o,s\rangle \;,\;\;\;\;\forall g\in
SO(3,2)^{\uparrow }
\end{equation}
Actually, the different elements $g\in SO(3,2)^{\uparrow }$ do not all
define the different states. The elements $g$ which change the ground state
only by a phase factor
$$
g|E_o,s\rangle =exp(i\alpha (g))|E_o,s\rangle
$$
form a subgroup commonly referred to as a {\it stability subgroup} of the
state $|E_o,s\rangle $.

From the definition (\ref{lw}) it follows that when $s=0$ the ground state
is invariant under the action of the maximal compact subgroup $SO(3)\times
SO(2)$ while for nonvanishing $s$ it is reduced to $SO(2)\times SO(2)$
generated by $L_{05}$ and $L_{12}$. Thus the set of CS is labelled by the
points of the quotient space ${\cal F}$ in the spinless case and $%
SO(3,2)^{\uparrow }/SO(2)\times SO(2)\sim {\cal F}\times S^2$, when $s\neq 0$%
.

In the next section we give an explicit realization for the CS in terms of
irreducible (spin-)tensor fields on AdS space.

\section{Coherent (spin-)tensor fields on AdS space}

Let us first consider the case of spinless particle on AdS space transformed
under representation $D(E_o,0)$. In the coordinate representation (\ref{5p})
this particle is described by the complex scalar field $\Phi (y)\equiv
\langle y^A|\Phi \rangle $, where $|\Phi \rangle \in D(E_o,0)$. In
accordance with (\ref{lw}), the ground state $\stackrel{\circ }{\Phi }%
(y)=\langle y^A|E_o,s\rangle $ arises as the solution for the following
system of equations:
\begin{equation}
\label{sfeq}
\begin{array}{c}
\langle y|L_{ij}|E_o,s\rangle =
{\cal L}_{ij}\stackrel{\circ }{\Phi }(y)=0 \\  \\
\langle y|L_{05}-E_o|E_o,s\rangle =(
{\cal L}_{05}-E_o)\stackrel{\circ }{\Phi }(y)=0 \\  \\
\langle y|E_i^{-}|E_o,s\rangle =({\cal L}_{i0}-i{\cal L}_{5i})\stackrel{%
\circ }{\Phi }(y)=0
\end{array}
\end{equation}
Up to a multiplicative constant, these equations have the unique solution
\begin{equation}
\label{sf}\stackrel{\circ }{\Phi }(y)=(y,\stackrel{\circ }{p}%
)=(y^0+iy^5)^{-E_o}=e^{-iE_ot}Y^{-E_o}
\end{equation}
Acting on this state by all the $SO(3,2)^{\uparrow }$ transformations one
gets the family of CS
\begin{equation}
\Phi (y)=(y,p)^{-E_o}\;,\;\;\;p_A=G_A{}^0+iG_A{}^5\;,\;\;\;G_A{}^B\in
SO(3,2)^{\uparrow }
\end{equation}
where by definition the complex vector $p_A$ obeys the conditions (\ref{p}),
(\ref{ppp}).

It is pertinent to note that the constructed basis of states will remain
overcomplete even though one restricts the possible values of $p^A$ onto the
`massive hyperboloid' setting $p_A=(i,\frac 1mp_\mu )$, where $\mu =0,1,2,3$
and $p_\mu $ is the real Lorentz vector. This vector may be regarded as the
four-vector of momentum of the particle in AdS space, in particular, the
state with minimal energy corresponds to the usual transition to the rest
reference system $p_\mu =(m,\vec 0)$. In this Lorenz-covariant
parametrization the CS for a spinless particle was originally considered by
Fronsdal \cite{II}.

Let as now consider the spinor field on AdS space $\Psi _a(y)\equiv\langle
y,a|\Psi \rangle $, $|\Psi \rangle \in D(E_o,1/2)$, transformed under the
action of $sp(4)$ by the law \footnote{%
All the details of the SO(3,2)-spinor formalism are collected in Appendix.}
\begin{equation}
\label{st}(L_{AB}\Psi )_a=i{\cal L}_{AB}\Psi _a+i(\Gamma _{AB})^b{}_a\Psi _b
\end{equation}
Then the corresponding system of differential equations for the ground looks
like
\begin{equation}
\label{seq}
\begin{array}{c}
\langle y,a|E_j^{-}|E_o,1/2\rangle =i(
{\cal L}_{0j}-i{\cal L}_{5j})\stackrel{\circ }{\Psi }_a+i((\Gamma
_{0j}-i\Gamma _{5j})\stackrel{\circ }{\Psi })_a=0 \\  \\
\langle y,a|L_{05}-E_o|E_o,1/2\rangle =(i
{\cal L}_{05}-E_o)\stackrel{\circ }{\Psi }_a+i(\Gamma _{05}\stackrel{\circ }{%
\Psi })_a=0 \\  \\
\langle y,a|S^{-}|E_o,1/2\rangle =i(
{\cal L}_{13}-i{\cal L}_{32})\stackrel{\circ }{\Psi }_a+i((\Gamma
_{13}-i\Gamma _{32})\stackrel{\circ }{\Psi })_a=0 \\  \\
\langle y,a|L_{12}|E_o,1/2\rangle =(i{\cal L}_{12}-\frac 12)\stackrel{\circ
}{\Psi }_a+i(\Gamma _{12}\stackrel{\circ }{\Psi })_a=0
\end{array}
\end{equation}
There are two linearly independent solutions to this system. In order to
present them in a transparent form, let us consider the constant $Sp(4)$%
-spinors $\zeta $ annihilated by the spinning part of the energy lowering
operators
\begin{equation}
\label{www}i(\Gamma _{0j}-i\Gamma _{5j})\zeta =0\;\Longleftrightarrow
\;(\Gamma _0-i\Gamma _5)\zeta =0
\end{equation}
Since the rank of the $SO(3)$-invariant matrix $(\Gamma _0-i\Gamma _5)$
equals $2$, the two basis solutions $\zeta ^{\pm }$ of (\ref{www}) may by
chosen to satisfy: $\Gamma _{12}\zeta ^{\pm }=\pm \zeta ^{\pm }$. As the
consequence of (\ref{www}) one can also find
\begin{equation}
\label{a2}i\Gamma _{05}\zeta ^{\pm }=-\frac 12\zeta ^{\pm }
\end{equation}
It is now evident that because $Sp(4)$-transformations (\ref{st}) act
separately on coordinates and indices one of the solutions to the system (%
\ref{seq}) may be obtained as the tensor product of a proper scalar field (%
\ref{sf}) and the spinor $\zeta ^{+}$, i.e.
\begin{equation}
\label{gs12}\stackrel{\circ }{\Psi }(y)=(y,\stackrel{\circ }{p})^{-E_o-\frac
12}\zeta ^{+}
\end{equation}
The second solution is immediately derived from (\ref{gs12}) if one observes
that the idempotent matrix $\widehat{y}=R^{-1}y_A\Gamma ^A$ defines the
invariant operator \footnote{%
In fact, this matrix may be used for the invariant definition of the right
(left) handed Weyl spinors $\chi _{+}$ $(\chi _{\_})$ on AdS space as
spinors obeying the conditions $\widehat{y}\chi _{\pm }=\pm \chi _{\pm }$ .
In this paper, however, only the Dirac spinors will be used.}
\begin{equation}
\label{y}[L_{AB},\widehat{y}]=0\ ,\qquad \widehat{y}^2=1
\end{equation}
such that
\begin{equation}
\label{sgs12}\stackrel{\circ }{\Psi }{}^{\prime }=\widehat{y}\stackrel{\circ
}{\Psi }=(y,\stackrel{\circ }{p})^{-E_o-\frac 12}\widehat{y}\zeta ^{+}
\end{equation}
automatically satisfies (\ref{seq}) if $\stackrel{\circ }{\Psi }$ does.

By applying to these states a general AdS-transformation one gets two
appropriate families of CS for describing a spin-1/2 particle
\begin{equation}
\label{ss12}{\Psi }(y)=(y,p)^{-E_o-\frac 12}\zeta \quad ,\qquad {\Psi }%
^{\prime }(y)=(y,p)^{-E_o-\frac 12}\widehat{y}\zeta ,
\end{equation}
where
\begin{equation}
\label{rtr}p_A\Gamma ^A\zeta =0\quad ,\qquad \zeta _a=G^b{}_a\zeta
_b^{+}\;,\;\;\;G^a{}_b\in Sp(4)
\end{equation}
In order to normalize these solutions we also put
\begin{equation}
\label{snorm}\bar \zeta \zeta =\sqrt{2}i
\end{equation}
The relationship between the spinor and vector representations of AdS group
is established by the standard relation
\begin{equation}
\label{svr}G^a\!_c(\Gamma _A)^c\!_dG^d\!_b=(\Gamma _B)^a\quad _bG^B\!_A
\end{equation}

Let us finally turn to the spin-1 case. For the sake of explicit $SO(3,2)$%
-covariance any tensor field on AdS space will be treated as a restriction
to the hyperboloid surface of the one defined on ${\bf R}^{3,2}$ and subject
to the $y$-transversality condition. For instance, a vector field $\Phi
_A(y) $ on ${\bf R}^{3,2}$ can be unambiguously restricted to the surface (%
\ref{p0}) if it obeys the condition
\begin{equation}
\label{yt}y^A\Phi _A(y)=0
\end{equation}
The action of the infinitesimal generators $L_{AB}$ of $SO(3,2)$ on this
field is given by
\begin{equation}
\label{vt}(L_{AB}\Phi )_C=i{\cal L}_{AB}\Phi _C+i\eta _{AC}\Phi _B-i\eta
_{BC}\Phi _A
\end{equation}
As is in the previous cases, the condition for the field $\Phi _A(y)$ to
realize the ground state of $D(E_o,1)$ leads to a set of equations, which
now have the form
\begin{equation}
\label{veq}
\begin{array}{c}
\langle y,A|E_j^{-}|E_o,1\rangle =i(
{\cal L}_{0j}-i{\cal L}_{5j})\stackrel{\circ }{\Phi }_A+i\eta _{0A}\stackrel{%
\circ }{\Phi }_j-i\eta _{jA}\stackrel{\circ }{\Phi }_0+ \\  \\
+\eta _{5A}
\stackrel{\circ }{\Phi }_j-\eta _{jA}\stackrel{\circ }{\Phi }_5=0 \\  \\
\langle y,A|L_{05}-E_o|E_o,1\rangle =(i
{\cal L}_{05}-E_o)\stackrel{\circ }{\Phi }_A+i\eta _{0A}\stackrel{\circ }{%
\Phi }_5-i\eta _{5A}\stackrel{\circ }{\Phi }_0=0 \\  \\
\langle y,A|S^{-}|E_o,1\rangle =i(
{\cal L}_{13}-i{\cal L}_{32})\stackrel{\circ }{\Phi }_A+i\eta _{1A}\stackrel{%
\circ }{\Phi }_3-i\eta _{3A}\stackrel{\circ }{\Phi }_1+ \\  \\
+\eta _{3A}
\stackrel{\circ }{\Phi }_2-\eta _{2A}\stackrel{\circ }{\Phi }_3=0 \\  \\
\langle y,A|L_{12}|E_o,1\rangle =(i{\cal L}_{12}-1)\stackrel{\circ }{\Phi }%
_A+i\eta _{1A}\stackrel{\circ }{\Phi }_2-i\eta _{2A}\stackrel{\circ }{\Phi }%
_1=0
\end{array}
\end{equation}
After straightforward calculations one finds that the system (\ref{yt}),(\ref
{veq}) possesses the unique (up to a constant) solution which may be written
as follows:

\begin{equation}
\begin{array}{c}
\stackrel{\circ }{\Phi }_A(y)=(y,\stackrel{\circ }{p})^{-E_o-1}h_A \\  \\
h_A=(y,\stackrel{\circ }{p})\stackrel{\circ }{q}_A-(y,\stackrel{\circ }{q})%
\stackrel{\circ }{p}_A\;\;,\;\;\;y^Ah_A\equiv 0
\end{array}
\end{equation}
where
\begin{equation}
\label{dq}\stackrel{\circ }{q}=(0,0,i,1,0)
\end{equation}
A coherent state of a general form is obtained by the replacement: $(%
\stackrel{\circ }{p},\stackrel{\circ }{q})\rightarrow (p,q)$, where by
definition the latter pair of vectors is constrained to satisfy
\begin{equation}
\label{hhh}q^2=(q,p)=(q,{\bar p})=0\;,\;\;\;(q,{\bar q})=2\;
\end{equation}
Rel. (\ref{svr}) makes it possible to reexpress the vector $q$ via $\zeta$
and ${p}$ as follows:
\begin{equation}
\label{lll}q^A=\frac 23{\bar p}_B\tilde \zeta \Gamma ^{BA}\zeta
\end{equation}

Now we are in a position to perform an explicit construction of CS for all
massive representations of AdS group in appropriate spaces of higher rank
(spin-)tensors. In order to do this, there is no need to write down and
solve a respective set of equations for a ground state. The ground states
for the higher-spin representations can be actually obtained by multiplying
the ones for spin-$1/2$ and spin-$1$ particles

\begin{equation}
\label{ground}\stackrel{\circ }{\Phi }_{A_1\cdots A_na_1\cdots a_m}=%
\stackrel{\circ }{\Phi }_{A_1}\cdots \stackrel{\circ }{\Phi }_{A_n}\stackrel{%
\circ }{\Psi }_{a_1}\cdots \stackrel{\circ }{\Psi }_{a_m}
\end{equation}
(For the sake of simplicity we use here only the first-type solution for the
spin-1/2 ground state. All subsequent results, however, can be readily
reformulated in terms of $\Psi ^{\prime }$.) In so doing, the weight of the
resulting ground-state is equal to the sum of weights of the multipliers.
From the group-theoretical viewpoint this procedure is identified with the
Young product of irreducible lowest-weight representations.

Thus, the most general family of CS obtained in such a way looks like
\begin{equation}
\label{cs}\Phi (p,\zeta |y)_{A_1\cdots A_na_1\cdots
a_m}=(y,p)^{-E_o-s}h_{A_1}\cdots h_{A_n}\zeta _{a_1}\cdots \zeta _{a_m}
\end{equation}
where $E_o$ is the minimal energy and%
$$
s=n+\frac 12m
$$
is the spin. The fields (\ref{cs}) are symmetric in both groups of indices,
traceless in its vector ones and possesses the $\Gamma $-transversality
condition:
\begin{equation}
\label{g}(\Gamma ^{A_1})^{a_1}{}_a\Phi _{A_1\cdots A_na_1\cdots a_m}=0
\end{equation}
Also, the following equations hold true: \refstepcounter{equation}\label{req}
$$
\left\{ i(\Gamma ^{AB})^a{}_{a_1}{\cal L}_{AB}+E_o+\frac 12m\right\} \Phi
_{A_1\cdots A_naa_2\cdots a_m}=0\eqno{(\theequation{}.a)}
$$
$$
\left\{ \frac 12{\cal L}^{AB}{\cal L}_{AB}-(E_o+\frac 12m)(E_o+\frac
12m-3)\right\} \Phi _{A_1\cdots A_na_1\cdots a_m}=0\eqno{(\theequation{}.b)}
$$
$$
y^{A_1}\Phi _{A_1\cdots A_na_1\cdots a_m}=0\quad ,\qquad \partial ^{A_1}\Phi
_{A_1\cdots A_na_1\cdots a_m}=0\eqno{(\theequation{}.c)}
$$
Eqs. (\ref{req}) constitute the full set of relativistic wave equations for
the irreducible massive fields on AdS space. Note that, when $m\neq 0,$ the
mass-shell condition (\ref{req}.b) follows from (\ref{req}.a). It should be
stressed that for a given quantum numbers $E_o,s$ only the first solution $%
\Psi $ of (\ref{seq}) obeys these equations. This observation allows to
remove the ambiguity in the choice of the ground state in the case $m\neq 0$%
. As a consequence, {\it the lowest weight state with given quantum numbers $%
E_o,s$ is unique in the space of solutions for} (\ref{req}). For some
special cases these equations present the AdS generalizations of the
well-known wave equations for the flat space-time. For example, for $m=0,$
we have an ordinary equations for massive tensor fields; in the case $m=1$
one may recognize the Rarita-Shwinger equations for half-integer spins;
setting $n=0$ we arrive at the Bargmann-Wigner equations, which for $m=1$ is
reduced to the equation for spin-$1/2$ particle in AdS space originally
proposed by Dirac \cite{Dirac}.

To conclude this Section, let us note that the relations (\ref{p}-\ref{ppp}%
), (\ref{rtr}, \ref{snorm}), (\ref{hhh}) imposed on the pair of parameters $%
p,\zeta $ (or $p,q$) define the ten-dimensional constrained surface being
isomorphic to the group manifold $Sp(4)$ (or $SO(3,2)^{\uparrow }$). The
phase transformations: $p\rightarrow e^{i\varphi }p,$ $\zeta \rightarrow
e^{i\psi }\zeta ,$ (or, respectively, $q\rightarrow e^{2i\psi }q$) preserve
this surface and the corresponding physical states (they acquire only a
phase factor). On the other hand, in Sec.2 it was argued that, for nonzero
spin, the set of CS is in one-to-one correspondence with the points of
eight-dimensional homogeneous space ${\cal F}\times S^2$. This implies that
the equations (\ref{p}-\ref{ppp}), (\ref{rtr}, \ref{snorm}), (\ref{hhh})
supplemented by the equivalence relations
\begin{equation}
\label{eqrel}\zeta \sim e^{i\psi }\zeta \quad ,\qquad q\sim e^{2i\psi
}q\quad ,\qquad \forall \psi \in {\bf R}
\end{equation}
present two covariant realizations for this space. Hereafter, we will refer
to ${\cal F}\times S^2$ as to the {\it dual space}.

\section{Unitarity and reproducing kernel}

In this section, we turn to the questions of irreducibility and unitarity of
the AdS-group representations in the spaces ${\cal H}_{E_o,s}$ of fields
spanned by the set of CS:
\begin{equation}
\label{h}\varphi _{A_1...A_na_1...a_m}(y)=\int_{{\cal F}\times S^2}\omega
(p,\zeta )\ \widetilde{\varphi }(p,\zeta )\Phi (p,\zeta
|y)_{A_1...A_na_1...a_m}\ \in {\cal H}_{E_o,s}
\end{equation}
where the invariant eight-form looks like
\begin{equation}
\label{form2}\omega =(d\overline{\zeta }_a\wedge d\zeta ^a)\wedge (d%
\overline{p}_A\wedge dp^A)^3
\end{equation}
and the coefficients $\widetilde{\varphi }(p,\varsigma )$ satisfy the
restrictions
\begin{equation}
\label{hc}\widetilde{\varphi}(e^{i\alpha }p,e^{i\beta }\zeta )=e^{i\alpha
(E_o+\frac 12m)-i\beta\frac{m}{2}}\widetilde{\varphi}(p,\zeta)
\end{equation}
for the integrand to be well-defined on the dual space ${\cal F}\times S^2$.
In fact, the last relation tells us that the coefficients $\widetilde{%
\varphi }$ belong to the space of special densities on ${\cal F}\times S^2$.
It should be noted, however,that this space is too large to be isomorphic to
${\cal H}_{E_o,s}$. The corresponding isomorphic subspace is extracted by a
projector to be specified further.

Hereafter we restrict our consideration to the case of integer energies $E_o$%
, then the quantum states of the particle will be described by single-valued
wave functions for the integer spins and double-valued for the halfinteger
ones. Upon this restriction the massive representations are characterized by
the inequality $E_o\geq s+2$.

To assign ${\cal H}_{E_o,s}$ with a structure of a Hilbert space carrying a
unitary representation of $SO(3,2)^{\uparrow }$ it is necessary to introduce
an invariant positive-definite inner product. In this case, the
irreducibility will follow directly from the uniqueness of the lowest weight
state. This may be seen as follows. The unitarity causes ${\cal H}_{E_o,s}$
to be a direct sum of irreducible subspaces, on the other hand, the spectrum
of the energy operator $L_{05}$ is bounded from below in ${\cal H}_{E_o,s}$
by construction. Hence, each irreducible component possesses a unique lowest
weight state and the number of components equals the number of ground states
which is just one in the case at hand. Indeed, if there exist another lowest
weight state $|E_o^{\prime },s^{\prime }\rangle \in {\cal H}_{E_o,s}$ then,
by construction, $E_o^{\prime }\geq E_o$. As the values of two Casimir
operators $C_2$ and $C_4$ (\ref{cas}) coincide for both weights $(E_o,s)$
and $(E_o^{\prime },s^{\prime })$ one gets the equations
\begin{equation}
C_2(E_o,s)=C_2(E_o^{\prime },s^{\prime })\quad ,\qquad
C_4(E_o,s)=C_4(E_o^{\prime },s^{\prime })
\end{equation}
The cases when these equalities takes place are
\begin{equation}
(E_o^{\prime },s^{\prime })=(E_o,s)\ ,\qquad (s+2,E_o-2)\ ,
\end{equation}
and those obtained from this two cases by the maps $E\rightarrow 3-E$ and $%
s\rightarrow 1-s$.

The first possibility corresponds only to the ground state as the
consequence of the uniqueness of the lowest weight state with the given
quantum numbers (see Sec. 2). In the case of massive representations, $%
E_o\geq s+2$, the rest possibilities are ruled out because $E_o^{\prime
}\leq E_o$.

The relevant inner product reads
\begin{equation}
\label{ip}\langle \varphi _1|\varphi _2\rangle =N^{-1}\int_{AdS}\Omega \,%
\overline{\varphi }_1(y)_{A_1\cdots A_na_1\cdots a_m}\varphi _2^{A_1\cdots
A_na_1\cdots a_m},
\end{equation}
where
\begin{equation}
\label{form1}\Omega =-\frac 1{24R}y^A\epsilon _{ABCDE}dy^B\wedge dy^C\wedge
dy^D\wedge dy^E
\end{equation}
is the invariant volume form on AdS space associated with the metric (\ref
{metric}). The inner product is manifestly AdS-invariant but its
positive-definiteness requires special study. With this is in mind consider
first the inner product of two CS. Substitution of (\ref{cs}) in (\ref{ip})
leads to the integral which converges only provided that $E_o+m/2>3/2$ and
the result is
\begin{equation}
\label{rk}
\begin{array}{c}
\displaystyle{\langle p^{\prime }\zeta ^{\prime }|p,\zeta \rangle =}\frac
\mu N{\frac{(\overline{\zeta }^{\prime }\zeta )^m}{(\overline{p}^{\prime
},p)^{E_o+s}}\left[ (\overline{p}^{\prime },p)(\overline{q}^{\prime },q)-(%
\overline{p}^{\prime },q)(\overline{q}^{\prime },p)\right] ^n} \\  \\
\displaystyle\mu {=\frac{(-2)^{E_o+\frac m2}2\pi ^2R}{E_o+s-1}B(\frac
12,E_o+\frac 12m-\frac 32)}
\end{array}
\end{equation}
It should be noted that due to the condition (\ref{ppp}) this expression is
well-defined for any two CS as $(\overline{p}^{\prime },p)$ can never come
to zero. In order to normalize the states we put $N=(-1)^{E_o}2^{n-E_o}\mu $%
. Then
\begin{equation}
\label{neq}0<|\langle p^{\prime }\zeta ^{\prime }|p,\zeta \rangle |\leq 1
\end{equation}
where the equality is reached only provided $p=p^{\prime }$ and $\zeta
=\zeta ^{\prime }$. We see that there are no two orthogonal states in the CS
set.

One can come to the expression (\ref{rk}) from the following line of
reasoning. First of all, it is sufficient to consider only the case when one
of the wavefunctions is the ground state, i.e. to evaluate the function $F(%
\overline{p},\overline{\zeta })=\langle p,\zeta |\stackrel{\circ }{p},\zeta
^{+}\rangle $, the general situation can then be restored by $%
SO(3,2)^{\uparrow }$-transformations. Exploiting the invariance property of
the inner product one may readily find that the function $F(\overline{p},%
\overline{\zeta })$ realizes the lowest weight state in the space of complex
densities on ${\cal F}\times S^2$ (\ref{hc}) or, what is the same, obeys the
following system of equations:
\begin{equation}
\label{dgs}
\begin{array}{c}
\langle p,\zeta |E_j^{-}|
\stackrel{\circ }{p},\zeta ^{+}\rangle =i({\cal L}_{0j}^{*}-i{\cal L}%
_{5j}^{*})F(\overline{p},\overline{\zeta })=0 \\  \\
\langle p,\zeta |L_{05}-E_o|
\stackrel{\circ }{p},\zeta ^{+}\rangle =(i{\cal L}_{05}^{*}-E_o)F(\overline{p%
},\overline{\zeta })=0 \\  \\
\langle p,\zeta |S^{-}|
\stackrel{\circ }{p},\zeta ^{+}\rangle =i({\cal L}_{13}^{*}-i{\cal L}%
_{32}^{*})F(\overline{p},\overline{\zeta })=0 \\  \\
\langle p,\zeta |L_{12}-s|\stackrel{\circ }{p},\zeta ^{+}\rangle =(i{\cal L}%
_{12}^{*}-s)F(\overline{p},\overline{\zeta })=0
\end{array}
\end{equation}
where
\begin{equation}
\label{dgen}{\cal L}_{AB}^{*}=\overline{p}_A\overline{\partial }_B-\overline{%
p}_B\overline{\partial }_A+(\Gamma _{AB})^b{}_a\overline{\zeta }_b\overline{%
\partial }^a
\end{equation}
Under the assumption of convergence of the integral (\ref{ip}), these
equations enable one to determine $F$ up to a multiplicative constant,
which, in turn, may be fixed by the direct evaluation of the norm of the
ground state $F(\stackrel{\circ }{\overline{p}}\overline{\zeta }{}^{+})$.

The inner product of CS (\ref{rk}) is said to define the {\it reproducing
kernel} of the CS family if it satisfies the equation
\begin{equation}
\label{pr}\int_{{\cal F}\times S^2}\omega \ (p,\zeta )\;\langle p^{\prime
},\zeta ^{\prime }|p,\zeta \rangle \langle p,\varsigma |p^{\prime \prime
},\zeta ^{\prime \prime }\rangle =c\langle p^{\prime },\zeta ^{\prime
}|p^{\prime \prime },\zeta ^{\prime \prime }\rangle
\end{equation}
for some finite constant $c$. The arguments similar to those that had been
used to establish the explicit expression for the inner product (\ref{rk})
show that the equation (\ref{pr}) always holds true and the only troublesome
moment is the convergence of the integral (\ref{pr}). It is equivalent to
the condition
\begin{equation}
\label{const}c=F(\stackrel{\circ }{\overline{p}},\overline{\zeta }%
^{+})^{-1}\int_{{\cal F}\times S^2}\omega \,|F(\overline{p}{,}\overline{%
\zeta }{)}|^2<\infty
\end{equation}
The function $c=c(E_o,s)$ is not easy to calculate explicitly. The rough
estimation shows that the integral (\ref{const}) converges when $E_o>s+5$
and diverges if $E_o<s+2$. In the former case, by making an appropriate
renormalization of the kernel the constant $c$ may be set to one and the
relation (\ref{pr}) turns to the conventional condition for the kernel of
the projector operator. This projector acts in the space of comlex densities
(\ref{hc}) and its image defines the AdS-invariant subspace, denoted ${\cal H%
}_{E_o,s}^{*},$ with the elements obeying the equation

\begin{equation}
\label{hspace}\widetilde{\varphi }(p^{\prime }\zeta ^{\prime })=c\int_{{\cal %
F}\times S^2}\omega (p,\zeta )\ \langle p^{\prime },\zeta ^{\prime }|p,\zeta
\rangle \widetilde{\varphi }(p,\zeta )
\end{equation}
This is just the above-mentioned condition that we imposed on $\widetilde{%
\varphi }$ in the definition of ${\cal H}_{E_o,s}$ (\ref{h}). The inverse
transformation formula

\begin{equation}
\label{invers}\widetilde{\varphi }(p,\zeta)=\frac 1c\int_{AdS}\Omega \,\bar
\Phi (p,\varsigma |y)_{A_1\cdots A_na_1\cdots a_m}\varphi (y)^{A_1\cdots
A_na_1\cdots a_m}
\end{equation}
establishes the isomorphism between ${\cal H}_{E_o,s}$ and ${\cal H}%
^*_{E_o,s}$. Thus, the formulas (\ref{h}), (\ref{hspace}), (\ref{invers})
may be considered as a generalized Fourier transform for the massive fields
over $D=4$ AdS space.

Now it is straightforward to check that under the condition (\ref{const})
the introduced above inner product actually defines a Hilbert space
structure in ${\cal H}_{E_o,s}$. Indeed,
\begin{equation}
\label{unit}
\begin{array}{c}
||\varphi ||^2=\int_{AdS}\Omega\;
\overline{\varphi }(y)_{A_1\cdots A_na_1\cdots a_m}\varphi ^{A_1\cdots
A_na_1\cdots a_m}= \\  \\
=\int_{{\cal F}\times S^2}\omega (p^{\prime },\zeta ^{\prime })\int_{{\cal F}%
\times S^2}\omega (p,\zeta )\stackrel{-}{\widetilde{\varphi }}(p^{\prime
},\zeta ^{\prime })\langle p^{\prime },\zeta ^{\prime }|p,\zeta \rangle
\widetilde{\varphi }(p,\zeta )=c\int_{{\cal F}\times S^2}\omega \ |%
\widetilde{\varphi }|^2
\end{array}
\end{equation}

As a result, we have proved that for $E_o>s+5$ the space ${\cal H}_{E_o,s}$
carries the unitary irredicible representation of $so(3,2)$ algebra, and
presented a generalized Fourier transform between ${\cal H}_{E_o,s}$ and $%
{\cal H}_{E_o,s}^{*}$, with explicit realization of corresponding Hilbert
space structures.

It is worth to remark that the reproducing kernel considered as the set of
functions $F_{p^{\prime },\zeta ^{\prime }}(p,\zeta )$ generates an
overcomlete basis of states in the space ${\cal H}_{E_o,s}^{*}$. In the
special case $m=0$ the CS of this type were previously considered in ref.%
\cite{BG} in the context of geometric quantization of a massive spinning
particle on AdS space. In that paper the dual space ${\cal F}\times S^2$ was
identified with the phase space of the massive spinning particle on AdS
space. Besides this, not long ago the lagrangian model of AdS massive
spinning particle was suggested, with the physical phase space being ${\cal F%
}\times S^2$ \cite{klss}. Therefore, the construction of this paper may be
thought of as the coordinate representation realization for the quantum
theory of the model of ref.\cite{klss}.

\section{Quasiclassical interpretation}

In the previous section we have presented the coordinate-type realization
for the quantum mechanics of the massive spinning particle in AdS space. The
use of the CS technique provides one with explicit expression for the wave
functions expanded via the overcomlete basis of states being invariant under
the group action. Also the CS are known as the states possessing a minimal
quantum uncertainty when one defines this notion in an invariant manner $%
\cite{Perelomov}$. As the result, many of properties of these states turn
out to be closely related to those for the classical theory. Very often this
provides the most natural way to assign the classical system to their
quantum analog.

Let us demonstrate that our CS are localized on the time-like geodesics of
the AdS space. For this end, consider the norm of an arbitrary coherent
state:
\begin{equation}
\label{rho}
\begin{array}{c}
\langle p,\zeta |p,\zeta \rangle =\int_{AdS}\Omega \;\rho (y)=1 \\
\\
\displaystyle{\rho (y)=}\frac{(-1)^{\frac m2}2^s}N{\frac{%
(|y^Ap_A|^2-|y^Aq_A|^2)^n}{|y^Bp_B|^{2(E_o+s)}}}
\end{array}
\end{equation}
Here $\rho $ is proved to be non-negative function and plays the role of a
probability density to find the particle at the space-time point $y$. Since
any two time-like geodesics as well as CS are obtained from each other by a
proper AdS-transformation, it is sufficient to establish the fact of
localizability only for the states with minimal energy. The latter are
associated with the following choice of the parameters:
\begin{equation}
\label{rrs}p=\stackrel{\circ }{p}=(i,1,0,0,0)\ ,\Rightarrow q=(0,0,\vec q)
\end{equation}
Then the expression (\ref{rho}) takes the form
\begin{equation}
\label{mrho}
\begin{array}{c}
\displaystyle{\rho _{E_o}(y)=\frac{(-1)^{\frac m2}2^s}N\frac{(R^2+(\vec
y,\vec n)^2)^n}{(R^2+\vec y^2)^{(E_o+s)}}}\quad , \\  \\
n^i=\frac 12\varepsilon ^i\!_{jk}q^j\overline{q}^k\quad ,\qquad \vec
n{}^2=1\quad .
\end{array}
\end{equation}
Here we have passed to the local coordinates $(\vec y,t)$ defined by (\ref
{1p}). Now it is clear that the function $\rho _{E_o}$ reaches the maximum
at the points of the geodesic line $\vec y=0$ which is naturally identified
with the world line of the massive particle in the state of the rest. But
this is exactly the geodesic associated to the vector $\stackrel{\circ }{p}$%
, as it was discussed in Sec.1. Thus, the complex vector $p_A$ labeles the
family of CS and, simultaneously, the set of geodesics on which the
corresponding coherent states are localized.

To assign a physical interpretation to the rest parameters let us introduce
the following real five-vector:
\begin{equation}
\label{spin}
\begin{array}{c}
N_A=\frac 14\varepsilon _{ABCDE}p^B
\overline{p}^Cq^D\overline{q}^E\ , \\  \\
p^AN_A=0\quad ,\qquad N^AN_A=1.
\end{array}
\end{equation}
The last relations imply that for a given point $p\in {\cal F}$ , there are
only two independent components of $N$ spanning a unit two-sphere $S^2$. The
pair $(p_A,N_B)$ represents the alternative covariant parametrization for
the dual space ${\cal F}\times S^2$. Let us show that the introduced vector $%
N$ has a direct physical interpretation as the vector collinear to the
vector of spin. Really, in the rest frame (\ref{rrs}) it takes the form $%
\stackrel{\circ }{N}=(0,0,\vec n)$, where $\vec n$ is defined as in (\ref
{mrho}). Identifying the spin of the particle $s$ with the absolute value of
the total angular momentum $\vec L=\{\frac 12\varepsilon _{ijk}L^{jk}\}$ in
the rest reference system one readily finds that

\begin{equation}
\label{sdir}(\vec n,\vec L)\Phi (\stackrel{\circ }{p},\vec{q}%
|y)_{A_1..A_na_1...a_m}=s\Phi (\stackrel{\circ }{p},\vec{q}%
|y)_{A_1..A_na_1...a_m}
\end{equation}
and thereby the unit vector $\vec n$ is oriented along the spin of the
particle.

As is seen from (\ref{mrho}), the probability density decreases monotonously
outward from the geodesic $\vec{y}=0$. In the limit $|\vec y|\rightarrow
\infty $ one finds the two types of asymptotical behavior: $\rho _{E_o}\sim |%
\vec{y}|^{2E_o+m}$ - along the spin direction, and $\rho _{E_o}\sim |\vec
y|^{2(E_o+s)}$ - in the orthogonal directions. So, the extent of decreasing
of $\rho _{E_o}$ increases with increasing $E_o$. The last fact agrees well
with usual physical notions: the greater the mass of the particle ($E_o$ in
our case) the smaller a region this particle is localized within.

\section{Conclusion}

The main results of this paper are the explicit realization for family of CS
in the spaces of (spin-)tensor fields on AdS space and the construction, in
the case $E_o>s+5$, of the Hilbert spaces ${\cal H}_{E_o,s}$ spanned by
these states carrying the UIR $D(E_o,s)$ of AdS group. Also, a generalized
Fourier transform is introduced which establishes the isomorphysm between $%
{\cal H}_{E_o,s}$ and the space ${\cal H}_{E_o,s}^{*}$ of special densities
on the dual space ${\cal F}\times S^2$ being the phase space of a massive
spinning particle. For the rest interval of energies, $s+1<E_o<s+5$, the
reproducing kernel technique used in the paper does not provide a direct
proof of the unitarity and irreducibility of AdS group representations in $%
H_{E_o,s}$. So, this case requires a special consideration.

The constructions of this paper could be relevant for the perturbative
quantum field theory calculations on the AdS space background, in
particular, the constructed family of CS may serve as a starting point for
the covariant derivation of the propagator for the higher spin fields. For
the low spins these propagators were derived in \cite{propagators}, but the
method used there becomes rather cumbersome for $s>2$ .

We have considered quasiclassical properties of CS and observed the
localizability of these states on the time-like geodesic lines of AdS space.
This has allowed us to assign a physical interpretation for the CS
parameters, namely, they have been shown to label the respective geodesic
line and spin direction.

\section{Acknowledgments}

The authors are thankful to S. M. Kuzenko, S. L. Lyakhovich and A. G.
Sibiryakov for fruitful discussions at various stages of the work. The work
is supported in part by the INTAS-RFBR grant No 95-829.

\section{Appendix. $SO(3,2)$-spinor formalism}

The generating elements of Clifford algebra $\Gamma _A$ are chosen to
satisfy
\begin{equation}
\Gamma _A\Gamma _B+\Gamma _B\Gamma _A=-2\eta _{AB}{\bf 1}\;\;,\;\;\;diag(%
\eta _{AB})=(--+++)
\end{equation}
This algebra has the unique nontrivial irreducible representation of complex
dimension $4$. Besides, there exists a Majorana representation in which all $%
\Gamma $-matrices are purely imaginary. Matrices $(\Gamma _A)^a{}_b$, $%
a,b=1,2,3,4$, allow a simple realization in terms of ordinary
four-dimensional Dirac $\gamma $ - matrices, namely, one may put: $\Gamma
_A=\gamma _A$, where $\gamma _5=i\gamma _0\gamma _1\gamma _2\gamma _3$.

16 matrices
\begin{equation}
\label{appendix1}{\bf 1}\;\;,\;\;\;\Gamma_A\;\;,\;\;\;\Gamma_{AB}=-\frac14[%
\Gamma_A,\Gamma_B]
\end{equation}
constitute the full linear basis in $Mat(4,C)$.

The charge and Hermitian conjugation automorphisms defined through the
matrices $C$ and $\Gamma $ are given by
\begin{equation}
\Gamma _A^t=C^{-1}\Gamma _AC\;,\;\;C^{\dagger }=C^{-1}\;,\;\;C^t=-C\;;
\end{equation}
\begin{equation}
\Gamma ^{\dagger }=-\Gamma \Gamma _A\Gamma ^{-1}\;,\;\;\Gamma ^{\dagger
}=-\Gamma \;,\;\;\Gamma ^2=-{\bf 1}\,.
\end{equation}
In representation, in which $\Gamma _0$ $\Gamma _5$ is Hermitean and $\Gamma
_i$ -- anti-Hermitean, $i=1,2,3$, one may put $\Gamma =\Gamma _0\Gamma _5$.
In the Majorana representation one has $C=\Gamma $.

The matrices $C^{ab}$ and $(C^{-1})_{ab}$ are used for the rising and
lowering of spinor indices.

There are following symmetry properties: $\Gamma _{AB}C$ are symmetric, $C$
and $\Gamma _AC$ are antisymmetric.

Matrices $\Gamma _{AB}$ realize the spinor representation of $so(3,2)\sim
sp(4)$. Exponential map
\begin{equation}
G=exp(\frac 12K^{AB}\Gamma _{AB})\;\;,\;\;\;K_{AB}=-K_{BA}\;\;,\;\;\;\bar
K_{AB}=K_{AB}
\end{equation}
defines the spinor representation of $Sp(4)$, which is the double covering
group of $SO(3,2)^{\uparrow }$, $SO(3,2)^{\uparrow }\sim Sp(4)/\pm {\bf 1}$

Dirac and charge conjugated spinors are defined as
\begin{equation}
\label{appp}\bar \psi =\psi ^{\dagger }\Gamma \;\;,\;\;\;\;\tilde \psi =\psi
^tC^{-1}
\end{equation}
The Majorana spinors are extracted by the condition $\bar \psi =\tilde \psi $%
.

The following useful formulas take place:\\

\noindent
1. Fierz identity for two spinors
\begin{equation}
\bar \vartheta \otimes \psi =\frac 14(\bar \vartheta \psi ){\bf 1}-\frac
14(\bar \vartheta \Gamma ^A\psi )\Gamma _A-\frac 12(\bar \vartheta \Gamma
^{AB}\psi )\Gamma _{AB}\,.
\end{equation}
2. Contractions of $\Gamma $ - matricies in vector indices
$$
\Gamma _{ab}^A(\Gamma _A)_{cd}=C_{ab}C_{cd}-2(C_{ac}C_{bd}-C_{bc}C_{ad})
$$
\begin{equation}
\Gamma _{ab}^{AB}(\Gamma _{AB})_{cd}=(C_{ac}C_{bd}+C_{bc}C_{ad})
\end{equation}
$$
\Gamma _{cm}^{AB}(\Gamma _B)_{nb}=\frac 12\left\{ (\Gamma
^A)_{cn}C_{mb}+(\Gamma ^A)_{cb}C_{nm}+(\Gamma ^A)_{mn}C_{cb}+(\Gamma
^A)_{mb}C_{nc}\right\}
$$
3. Contractions of $\Gamma $ - matrices in spinor indices
\begin{equation}
Tr(\Gamma _A\Gamma _B)=-4\eta _{AB}
\end{equation}
\begin{equation}
Tr(\Gamma _{AB}\Gamma _{CD})=\eta _{BC}\eta _{AD}-\eta _{AC}\eta _{BD}
\end{equation}

\end{document}